\documentclass[universe,article,accept,pdftex,oneauthor]{Definitions/mdpi} 
\usepackage{slashed}
\newcommand{\tmop}[1]{\ensuremath{\operatorname{#1}}}
\firstpage{1} 
\pubvolume{1}
\issuenum{1}
\articlenumber{0}
\pubyear{2024}
\copyrightyear{2024}
\externaleditor{Academic Editor: Firstname \linebreak Lastname}
\datereceived{2 July 2024} 
\daterevised{20 August 2024} 
\dateaccepted{21 August 2024} 
\datepublished{ } 
\hreflink{https://doi.org/} 

\Title{Infrared Regularization of Very Special Relativity Models}

\TitleCitation{Infrared Regularization of Very Special Relativity Models}

\Author{Jorge Alfaro \orcidA{}} 

\AuthorNames{Jorge Alfaro}

\AuthorCitation{Alfaro, J.} 

\address[1]{Facultad de F\' isica, Pontificia Universidad Cat\'olica de Chile, Santiago 7820436, Chile; jalfaro@uc.cl} 

\abstract{We extend the $\tmop{Sim}(2)$ invariant infrared regularization of Very Special Relativity models, that we have proposed recently, to include $\gamma_5$ Dirac matrix. Then, we solve the Very Special Relativity Schwinger model, find the chiral anomaly, and clarify its meaning in the new context. In addition, we show that the triangle anomaly in four space-time dimensions agrees with the same object in standard quantum electrodynamics. Finally, we apply the infrared regularization to compute the large N limit of the Very Special Relativity Gross--Neveu model.
}


\begin{document}

\section{Introduction}\label{sec:1}

One of the most puzzling problems of high energy physics is the origin of
neutrino masses.
In the Standard Model (SM) of weak and electromagnetic interactions, neutrinos
are massless and have left-handed chirality. However, we know that some of the
three species of neutrinos must have a mass, in order to explain the phenomenon of
neutrino oscillations~\cite{Langacker}.
In a Lorentz invariant theory, SM neutrinos acquire a mass, assuming the
existence of heavy right-handed neutrinos, as in the seesaw mechanism~\cite{mohapatra}.
To obtain neutrino masses, an alternative is to break Lorentz symmetry. 

Special Relativity has been tested and remains valid at very high energy~\cite{PA}. However, its violation has been proposed as a possible road to new physics. Lorentz invariance
can be violated in models of quantum gravity~\cite{AC,JLM}, in constant background fields derived from the spontaneous symmetry breaking of the Lorentz symmetry 
of a still unknown, more basic theory~\cite{MP,AA,CK}, and in Very Special Relativity (VSR)~\cite{CG1}.

VSR proposes that the fundamental symmetry of nature is not Lorentz's six
parameters group, but a subgroup of it, such as $T(2), E(2), HOM(2)$ or $\tmop{Sim}(2)$. The most interesting of these turns out to be the four parameters $\tmop{Sim}(2)$
group. In $\tmop{Sim}(2)$, there are no more invariant tensors besides the ones
that are invariant under the whole Lorentz group, so all the classical effects of
Special Relativity also hold true in VSR with $\tmop{Sim}(2)$ symmetry.
$\tmop{Sim}(2)$ is characterized by leaving a null vector $n_{\mu}$
invariant, except by a multiplicative factor, $n_{\mu} \rightarrow e^{\phi}
n_{\mu}$, so ratios of scalars such as $\frac{n.p_1}{n.p_2}$ are $\tmop{Sim}(2)$ invariants but violate Lorentz. Here, $p_{i \mu}$ are vectors in
space--time.
$\tmop{Sim}(2)$ allows for an additional term in the Dirac equation so that chiral
particles acquire a mass~\cite{CG2}. VSR has been generalized to consider supersymmetry
~\cite{CZ,Vo}, curved spaces~\cite{GG,Mu}, noncommutativity~\cite{ST,Das}, cosmological constant~\cite{AV}, dark
matter~\cite{AH}, cosmology~\cite{CL}, Abelian gauge fields~\cite{Cheon} and non-Abelian gauge fields~\cite{AR}.

Using this idea, we built the VSR SM~\cite{ja1}. It has the same symmetry and particles
as the SM, $\tmop{SU} (2)_L \times U (1)_R$, but neutrinos obtain a VSR mass.
Neutrino oscillations appear naturally and new processes beyond the SM are
predicted, such as $\mu \rightarrow e + \gamma$.

Loop corrections were difficult to compute due to the absence of a
$\tmop{Sim}(2)$ invariant regularization that preserves gauge invariance.
Recently~\cite{jaren}, we have introduced a $\tmop{Sim}(2)$ infrared regularization
inspired by physical considerations when we study the non-relativistic
potential in VSR models with VSR massive gauge particles. Without this
specific regularization, the non-relativistic potential (Coulomb potential)
exhibits singularities at certain angles for all radial distances, creating infinite electric forces there, which are ruled out by experiments. We have tested the new infrared
regularization, calculating the one loop renormalization of VSR quantum
electrodynamics with a gauge invariant photon mass $m_\gamma$. A remarkable
contribution to the anomalous magnetic moment of the electron appears,
depending on the electron neutrino and photon mass. It fits very well between
the bounds of the most recent measurements.

The infrared regularization of~\cite{jaren} has been applied to compute the scattering of light by light in VSR QED with a VSR photon mass~\cite{jalbl}.
Gauge invariance and $\tmop{Sim}(2)$
 symmetry are preserved. We obtain an analytic result for the total cross-section. Tiny new contributions show up due to the photon mass.
They are anisotropic and can be tested at cosmological scales.

In this paper, we extend the infrared regulator to include $\gamma_5$ Dirac matrix. We use it to compute the axial anomaly in two and four dimensions. We compare these calculations with the results obtained in~\cite{AS2,jaanom}. By computing the fermion mass contribution to the divergence of the chiral current, we are able to clarify the meaning of both the previous calculations of the chiral anomaly.
Finally, we apply the infrared regulator to solve the Gross--Neveu (GN) model with a VSR fermion mass in the large $N$ limit. Now two phases appear: in one of them, the chiral symmetry is broken, as usual; in the other phase, the chiral symmetry is unbroken.

The paper is written as follows: Section~\ref{sec:2} reviews the infrared regularization of~\cite{jaren}. Section~\ref{sec:3} defines the Lagrangian of quantum electrodynamics with VSR masses for the electron and photon. Section~\ref{sec:4} discusses the VSR Schwinger model. It contains the computation of the self energy of the photon, the two dimensional axial anomaly, and the fermion mass contribution to the divergence of the axial current. In Section~\ref{sec:5}, we use the infrared regulator to compute the axial anomaly in four dimensions. In Section~\ref{sec:6}, we solve the GN model with a VSR mass in the large $N$ limit. Section~\ref{sec:7} is devoted to the conclusions and discussions.

\section{\boldmath{$\tmop{Sim}(2)$} Invariant Regularization}\label{sec:2}
Here we review the regularization procedure introduced in~\cite{jaren}.
Consider an arbitrary function $g$ and compute

\[ \int \frac{d^d q}{(2 \pi)^d} g (q^2, q.x) \frac{1}{(n.q)^a}.\]

The Mandelstam--Leibbrandt (M-L) prescription~\cite{ML,Lei} using the method of~\cite{AML}, which implies
\[ \int \frac{d^d q}{(2 \pi)^d} g (q^2, q.x) \frac{1}{(n.q)^a} = (\bar{n}
.x)^a f (x.x, n.x \bar{n} .x) \]
for a unique $f (x.x, n.x \bar{n} .x)$, under the conditions:
\begin{enumerate}
	\item $n.n = 0 = \bar{n} . \bar{n}$, $n. \bar{n} = 1$
	
	\item Scale invariance under $n_{\mu} \rightarrow \lambda n_{\mu},
	\bar{n}_{\mu} \rightarrow \lambda^{- 1} \bar{n}_{\mu}$.
	
	\item $f (x.x, n.x \bar{n}.x)$ must be regular at $n.x \bar{n}.x = 0$.
\end{enumerate}
$x_{\mu}$ is an arbitrary vector.

To calculate the limit $\bar{n}_{\mu} \rightarrow 0$, write $\bar{n}_{\mu} = \rho
\bar{n}_{\mu}^{(0)}, n_{\mu} = \rho^{- 1} n_{\mu}^{(0)}$, with

$\bar{n}_{\mu}^{(0)}, n_{\mu}^{(0)}$ satisfying condition 1. Then, premise 1
is valid for all $\rho$.

We define $\bar{n}_{\mu} \rightarrow 0$ by the limit $\rho \rightarrow 0$.

We obtain: $\lim_{\rho \rightarrow 0} \rho^a (\bar{n}^{(0)} .x)^a f (x.x, n^{(0)}
.x \bar{n}^{(0)} .x) = 0$.

Thus:

\vspace{-6pt}
\[ \int \frac{d^d q}{(2 \pi)^d} g (q^2, q.x) \frac{1}{(n.q)^a} = 0, a > 0. \]

Clearly, this result applies to loop integrals of the kind
{~\cite{AML}},
\small{
\begin{eqnarray}
	\int dp \frac{1}{[p^2 + 2 p.q - m^2 + i \varepsilon]^a}  \frac{1}{(n \cdot
		p)^b} = (- 1)^{a + b} i (\pi)^{\omega}  (- 2)^b \frac{\Gamma (a + b -
		\omega)}{\Gamma (a) \Gamma (b)}  (\bar{n} \cdot q)^b &  &  \nonumber\\
	\int_0^1 dtt^{b - 1}  \frac{1}{(m^2 + q^2 - 2 n \cdot q \bar{n} \cdot qt - i
		\varepsilon)^{a + b - \omega}}, & \omega = d / 2 &  \label{basic}
\end{eqnarray}}

That is, the $\tmop{Sim}(2)$ limit is:
\[ \int dp \frac{1}{[p^2 + 2 p.q - m^2 + i \varepsilon]^a}  \frac{1}{(n \cdot
	p)^b} = 0, b > 0, q_{\mu} \mbox{arbitrary.}\]

Computing derivatives in $q_{\mu}$, we obtain:
\[ \int dp \frac{1}{[p^2 + 2 p.q - m^2 + i \varepsilon]^a}  \frac{p_{\alpha_1}
	\ldots .p_{\alpha_n}}{(n \cdot p)^b} = 0, b > 0, q_{\mu} \mbox{arbitrary.}
\]

Summarizing, the $\tmop{Sim}(2)$ invariant regularization of any integral over
$p_{\mu}$, containing $\frac{1}{n.p}$ to any positive power must be put
to zero.
Clearly, this operation respects gauge invariance and $\tmop{Sim}(2)$ invariance.	
But how should we continue if $\gamma$ matrices are present? 

Let us consider an example:
\begin{eqnarray*}
	\int \frac{dp}{p^2 - m^2} \frac{\not{n}  \not{p}  \not{n}}{n.p} =
	\int \frac{dp}{p^2 - m^2} \frac{\left( 2 n.p - \not{p \not{n}}
		\right)  \not{n}}{n.p} = 2 \int \frac{dp}{p^2 - m^2} &  & 
\end{eqnarray*}

We can compute $p$ integral first, using M-L:
\begin{eqnarray*}
	\int \frac{dp}{p^2 - m^2} \frac{\not{n}  \not{p}  \not{n}}{n.p} =
	\not{n} \gamma_{\mu}  \not{n} \int \frac{dp}{p^2 - m^2}
	\frac{p_{\mu}}{n.p} = \not{n} \gamma_{\mu}  \not{n} \int
	\frac{dp}{p^2 - m^2} \bar{n}_{\mu} &  & 
\end{eqnarray*}

The naive limit will give zero, but if we displace $\not{n}$ to the right (or left) $\not{n}  \not{\bar{n}}  \not{n} = 2$,
we obtain the same result as before.

However, assume that $\not{n}$ was previously on the right. Consider:
\begin{eqnarray*}
	\int \frac{dp}{p^2 - m^2} \frac{ \not{p}  \not{n}}{n.p} = \int
	\frac{dp}{p^2 - m^2}  \not{\bar{n}}  \not{n} \rightarrow ? &  & 
\end{eqnarray*}

Prescription:
We displace all $\not{n}$ to the right, collect all $n.(p+Q)$ produced by this motion and use them to cancel as many $n.(p+Q)$ in the denominator as possible. At the end, all remaining  $(n.(p+Q))^{-a},a>0$ are substituted by zero. Here, $Q_\mu$ means any vector different from $p_\mu$ (the integration variable), the zero vector included. Observe that $\frac{n.p}{n.(p+q)}=1$ because $\frac{n.p}{n.(p+q)}=1-\frac{n.q}{n.(p+q)}$ and the second term vanishes in the last step of the process.

That is:
\begin{eqnarray*}
	\int \frac{dp}{p^2 - m^2} \frac{ \not{p}  \not{n}}{n.p} =0 
\end{eqnarray*}

and
\begin{eqnarray*}
	\int \frac{dp}{p^2 - m^2} \frac{\not{n}  \not{p}  \not{n}}{n.p} =
	2 \int \frac{dp}{p^2 - m^2} &  & 
\end{eqnarray*}

This method works for the following reasons. We want to put $\bar{n}_\mu=0$ to recover $\tmop{Sim}(2)$ invariance. But we do not want to lose gauge invariance. Gauge invariance is reflected in Ward identities that the Feynman graphs must fulfil. If we order all graphs in a ``canonical form'', such as ensuring all $\not{n}$ are to the right in all monomials (only one $\not{n}$ remains, since $\not{n}.\not{n}=0$ ), the Ward identities that generally involve products with external momenta will be fulfilled for arbitrary values of $n_\mu$ and $\bar{n}_\mu$. (To prove the Ward identity we do not use the restrictions $n.n=\bar{n}.\bar{n}=0,n.\bar{n}=1$  anymore when all $\not{n}$ are to the right of all $\not{\bar{n}}$). Then, after evaluating $\bar{n}_\mu=0$, the Ward identity will still be valid in the surviving set of integrals defining the graphs. This surviving ensemble determines the $\tmop{Sim}(2)$ invariant gauge theory. 

This method has some looseness. Equally, we could use the criterion of ordering all $\not{n}$ to the left.

In the calculations in VSR QED, we have verified whether this looseness in the method produces ambiguities or not. We did not find any.

In Appendix~\ref{App:b}, we introduce the method of traces to obtain the $\bar{n}_\mu=0$ limit. Using the trace method, it is apparent that the Ward identities are fulfilled for the $\bar{n}_\mu=0$ ensemble. In VSR QED, the trace method produces the same outcome as the one presented in this chapter.

In the following chapters, we will see how to implement the  $\bar{n}_{\mu} \rightarrow 0$ limit when $\gamma_5$ is present.

\section{The Model}\label{sec:3}
The leptonic sector of VSRSM consists of three $SU (2)$ doublets $L_{a} =
\left( \begin{array}{c}
	\nu^{0}_{aL}\\
	e^{0}_{aL}
\end{array} \right)$, where $\nu^{0}_{aL} = \frac{1}{2}  (1- \gamma_{5} )
\nu^{0}_{a}$ and $e^{0}_{aL} = \frac{1}{2}  (1- \gamma_{5} ) e^{0}_{a}$, and
three $SU (2)$ singlet $R_{a} =e^{0}_{aR} = \frac{1}{2}  (1+ \gamma_{5} )
e^{0}_{n}$. We accept that there is no right-handed neutrino. The index $a$
classifies the different families and the index $0$ means that the fermionic
fields are the physical fields before spontaneous symmetry breaking.

In this work, we restrict ourselves to the electron family. It consists of the $e_L$ (the left-handed electron) and $\nu_e$(the electron's neutrino), forming a doublet of $SU(2)_L$, and $e_R$(the right-handed electron), which is a $SU(2)_L$ singlet. In order to respect the $SU(2)_L$ symmetry, we introduce a VSR mass $m$ for the doublet. Then,
$m$ is the VSR mass of both electron and neutrino. 
After spontaneous symmetry breaking (SSB), the electron acquires a mass term
$M= \frac{G_{e} v}{\sqrt{2}}$, where $G_{e}$ is the electron Yukawa coupling
and $v$ is the VEV of the Higgs. Please see Equation (52) of {~\cite{ja1}}. The electron mass is
$M_e = \sqrt{M^2 + m^2}$. The neutrino mass is not affected by SSB:$M_{\nu_e}= m$.

Restricting the VSRSM after SSB to photon($A_{\mu}$) and electron($\psi$) alone, neglecting the terms in the VSRSM that contain the neutrino and the gauge bosons $Z_0,W^{\pm}$, we obtain the VSR
QED action. Furthermore, we add a VSR mass $m_{\gamma}$ for the photon. We use the Feynman gauge.
\begin{eqnarray}
	\mathcal{L} = \bar{\psi}  \left( i \left( \not{D} + \frac{1}{2} \not{n} m^2
	(n \cdot D)^{- 1} \right) - M \right) \psi - \frac{1}{4} F_{\mu \nu} F^{\mu
		\nu} &  & \nonumber\\
	\hspace{-1cm}- \frac{1}{2} m_{\gamma}^2  (n^{\alpha} F_{\mu \alpha}) \frac{1}{(n \cdot
		\partial)^2}  (n_{\beta} F^{\mu \beta}) - \frac{(\partial_{\mu}
		A_{\mu})^2}{4} 
\end{eqnarray}
where $D_{\mu} = \partial_{\mu} - ieA_{\mu}, F_{\mu \nu} = \partial_{\mu} A_{\nu} -
\partial_{\nu} A_{\mu}$ and $n.n = 0$.

This Lagrangian (without the gauge-fixing term) is the gauge invariant under the
usual gauge transformations: $\delta A_{\mu} (x) = \partial_{\mu} \Lambda (x)$.
This is a crucial property of a VSR mass for the photon. It conserves
gauge invariance, as opposed to a Lorentz invariant mass for the photon, which destroys gauge
invariance. 

The Feynman rules are written in Appendix~\ref{App:a}.

The vector current (electric charge conservation) is:
\begin{equation}
	j^{\mu} = \bar{\psi}   \gamma^{\mu}   \psi + \frac{1}{2} m^{2} \left(
	\frac{1}{n  \cdot D^{\dag}} \bar{\psi} \right) \slashed{n} n^{\mu} \left(  
	\frac{1}{n  \cdot D}   \psi \right)\label{vc}
\end{equation}

The axial vector current is:
\begin{equation}
j^{\mu 5} =  \bar{\psi}   \gamma^{\mu}   \gamma^{5} \psi + \frac{1}{2}m^{2} \left( \frac{1}{n  \cdot D^{\dag}} \bar{\psi} \right) \slashed{n} n^{\mu}
\gamma^{5} \left(   \frac{1}{n  \cdot D}   \psi \right)\label{avc}
\end{equation}

At the classical level, 
(\ref{vc}) is conserved for any $M,m$.
(\ref{avc}) is conserved if $M=0$.~\cite{AS2}

We are interested in computing expectation values of these currents.

\section{VSR Schwinger Model}\label{sec:4}

A word of caution: In this section, we work in a 1 + 1 space--time, so the Lorentz group is a one parameter group. As we discussed in~\cite{AS2}, the two dimensional Lorentz group acts as a scale transformation on the null vector $n_\mu=(1,1)$. So, we can introduce VSR-like mass terms for the fermions. When we refer to $\tmop{Sim}(2)$ symmetry or regulator in two dimensions, we mean the scale transformation mentioned above.

For a previous discussion of this model, please see~\cite{AS2}.
\subsection{Photon Self Energy}
Let us compute the photon self-energy. 
	It is given by two graphs:
	
	\begin{figure}[H]
		\includegraphics[width=0.4\textwidth]{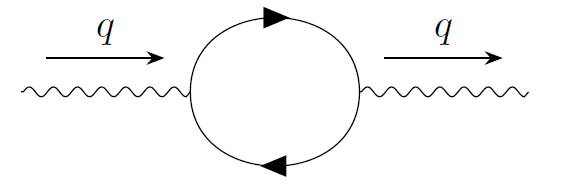}
		\includegraphics[width=0.4\textwidth]{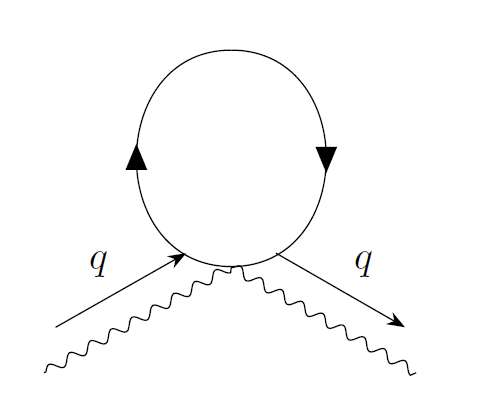}	
		\caption{Vacuum polarization one loop graphs in two dimensional VSR QED.}
		\label{Fig:vp}
	\end{figure}
	
	\vspace{-12pt}
	\begin{eqnarray}
		i \Pi_{1 \mu \nu} =-e^2 \int dp Tr [  ( \gamma_{\mu} +
		  \frac{n_{\mu}   \not{n}  m^{2}}{2n.(p+q)n.p} )  \frac{
			\left( \not{p} +M-  \frac{m^2\not{n}}{2n \cdot p} \right)}{p^{2}
			-M_{e}^{2} +i \varepsilon} &  & \nonumber\\
		( \gamma_{\nu} +  \frac{n_{\nu}   \not{n} 
			m^{2}}{2n.(p+q)n.p}  ) \frac{\left( \not{p} + \not{q} +M-
			 \frac{m^2\not{n}}{2n \cdot (p+q)} \right)}{(p+q)^{2} -M_{e}^{2}
			+i \varepsilon}   ]
	\end{eqnarray}
	\begin{eqnarray}
		i \Pi_{2 \mu \nu} =-e^{2 }  n_{\mu} n_{\nu}  \int dp \frac{1}{n.p n. (p+q)
			n. (p-q)} Tr (  \not{n} m^{2}  \frac{ \left( \not{p} +M-
			 \frac{m^2\not{n}}{2n \cdot p} \right)}{p^{2} -M_{e}^{2} 
			+i \varepsilon} )  &  & 
	\end{eqnarray}
	
	We use the  $\bar{n}_{\mu} \rightarrow 0$ limit to evaluate the diagrams. 
	
	The second graph of Figure 1 vanishes because there are 3 $n.p$ in the denominator and at most 1 $n.p$ in the numerator of the integrand. So, the $\bar{n}_{\mu} \rightarrow 0$ limit is zero. The count of $n.p$ is determined by $\tmop{Sim}(2)$ symmetry. The second graph has 2 $n_\mu$ outside the integral, so there must be 2 $n.p$ in the denominator ($n.q$ is not factored out the integral).
	The first graph of Figure 1 gives:
	\begin{eqnarray}
		i  \Pi_{1\mu \nu} =4 e^2\int \frac{d^d p}{(2\pi)^d}   \frac{-2p_{\mu} p_{\nu} -p_{\mu} q_{\nu}
			-p_{\nu} q_{\mu} - \eta_{\mu \nu} ( M_{e}^{2} -p^{2} -p.q )}{( p^{2}
			-M_{e}^{2} +i \varepsilon ) ( ( p+q )^{2} -M_{e}^{2} +i \varepsilon )} &  & 
	\end{eqnarray}
	
	Observe that some terms proportional to $m^{2}$ remain. They are produced by terms
	created by the trace of the form: $m^{2} n.p,m^{2} n. ( p+q )$. These terms
	balance the $\frac{1}{n.p},\frac{1}{n.(p+q)}$, so that after computing the $\tmop{Sim}(2)$ limit, they remain. These are precisely the factors we need to put the final
	result entirely in terms of the physical electron mass, $M_{e}^{2} =M^{2} +m^{2}$, which is expected from unitarity.	
	This is the  QED result, with the electron mass $M_{e}=\sqrt{M^2+m^2}$.
	
	Using dimensional regularization, we obtain:
\begin{equation}
i  \Pi_{1\mu \nu} = - 2 e^2 i \frac{\tmop{tr} (1)}{(4 \pi)^{d / 2}} \Gamma \left( 2 -
\frac{d}{2} \right) \int_0^1 \frac{d x}{(- q^2 x (1 - x) + M_e^2)^{2 -
		\frac{d}{2}}} x (1 - x) (q^2 \eta_{\mu \nu} - q_{\mu} q_{\nu}) 	
\end{equation}

Put:
\[ \tmop{tr} (1) = 2, d = 2 \]
\begin{eqnarray*}
	i \Pi_{\mu \nu} = - \frac{e^2}{\pi} i (q^2 \eta_{\mu \nu} - q_{\mu} q_{\nu})
	\int_0^1 \frac{d x}{(- q^2 x (1 - x) + M_e^2)} x (1 - x) &  & 
\end{eqnarray*}

Notice that if the fermion has a VSR mass  $m \neq 0$,$M=0$, the photon remains massless. Only when $m=M=0$, the photon obtains a mass $m_\gamma^2=\frac{e^2}{\pi}$.

Using  Equation (19.15) of~\cite{Peskin}, we obtain the vector current in the presence of an external electromagnetic field $A_\mu$:
\vspace{-3pt}
\begin{eqnarray*}
<j^\mu(q)>=	\int d^2 x < j^{\mu} (x) > e^{i q.x} = \frac{i}{e} (i \Pi^{\mu \nu} (q)
	A_{\nu} (q)) &  & 
\end{eqnarray*}

That is:
\vspace{-3pt}
\begin{equation}
<j^\mu(q)>= \frac{e}{\pi}  (q^2 \eta^{\mu \nu} - q^{\mu} q^{\nu})A_\nu(q)
\int_0^1 \frac{d x}{(- q^2 x (1 - x) + M_e^2)} x (1 - x)  	
\end{equation}

It satisfies the conservation law: $q_\mu<j^\mu(q)>=0$.

\subsection{Two-Dimensional Axial Anomaly}
In this case, we have to compute the expectation value of the axial vector
current in a background field $A_{\nu}$. We use the convention of~\cite{Peskin}, $\epsilon^{01}=+1$. 
\begin{equation}
	<j^{5\nu} ( q ) > =\int d^{2}x <j^{5\nu}( x ) >e^{i q x} = ( -i e
	)^{-1} i  \Pi^{5\mu \nu} ( q ) A_{\mu}
\end{equation}

The contribution to the two-dimensional axial vector current in VSR electrodynamics is
given by the two graphs (Figures \ref{fig:2d1} and \ref{fig:2d2}):
\vspace{-9pt}
\begin{figure}[H]
	\includegraphics[scale=1]{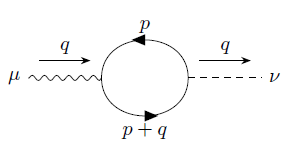}
	\caption{$i\Pi^{15\mu\nu}$ contribution to the two-dimensional axial vector current in VSR QED.}
	\label{fig:2d1}
\end{figure}
\vspace{-9pt}

\begin{figure}[H]
	\includegraphics[scale=1]{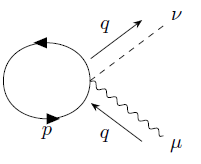}
	\caption{$i\Pi^{25\mu\nu}$ contribution to the two-dimensional axial vector current in VSR QED.}
	\label{fig:2d2}
\end{figure}
\vspace{-12pt}

\begin{eqnarray}
	i  \Pi^{ 1 5 \mu \nu} =- ( -i e )^{2} 
	\int d p  \tmop{Tr} \{
	\left[ \gamma^{\mu} + \frac{1}{2} n^{\mu} \left( \slashed{n} \right) m^{2}   (n. ( p+q ) )^{-1} ( n.p )^{-1} \right] \nonumber \\
	\frac{i \left( \slashed{p} +M-\frac{m^{2}}{2}  \frac{\slashed{n}}{n \cdot p} \right)}{p^{2} -M^{2} -m^{2} +i\varepsilon}
	\left[ \gamma^{\nu} + \frac{1}{2} n^{\nu} \left( \slashed{n}
	\right) m^{2}   ( n. ( p+q ) )^{-1} ( n.p )^{-1} \right]
	\gamma^{5} 
	\frac{i \left( \left( \slashed{p} + \slashed{q} \right) +M- \frac{m^{2}}{2} 
		\frac{\slashed{n}}{n \cdot ( p+q )} \right)}{( p+q )^{2} -M_e^{2}  +i
		\varepsilon} \} \label{2d1}
\end{eqnarray}

\vspace{-6pt}
\begin{eqnarray}
	i  \Pi^{2 5\mu \nu} = ( -1 ) ( i e )^{2} n^{\mu} n^{\nu} i \int  d p (
	n.p )^{-1} ( n.p )^{-1} [ ( n. ( q+p ) )^{-1} + ( n. ( -q+p ) )^{-1} ]\nonumber\\
	\tmop{Tr} \{\frac{1}{2} \slashed{n} m^{2}  \frac{i \left( \slashed{p} +M-
		\frac{m^{2}}{2}  \frac{\slashed{n}}{n \cdot p} \right)}{p^{2} -M_e^{2}  +i
		\varepsilon} \gamma^{5}\} &  &  \label{2d2}
\end{eqnarray}

Notice that this is a new vertex, one photon line, and one axial vector
line (current insertion). Please see Figure~\ref{Fig: Fey rulesA}.

$\Pi^{2 5\mu \nu}$ is zero in the $\tmop{Sim}(2)$ limit. In fact, there are 3 $n.p$ in the denominator and at most 1 $n.p$ in the numerator of the integrand, where $p_\mu$ is the integration variable. According to the rules of Section~\ref{sec:2}, the integral vanishes when $\bar{n}_{\mu} \rightarrow 0$.
Now, we compute the $\tmop{Sim}(2)$ limit of $\Pi^{15 \mu \nu}$.

We have to study the behavior of the
$\bar{n}_{\mu} \rightarrow 0$ limit in the presence of $\gamma_5$. It is
straightforward to do so. We must use the standard
definition of
$\gamma_5$ in the corresponding dimension from the beginning, $\gamma_5 = i \gamma_0 \gamma_1$,
in two dimensions; $\gamma_5 = i \gamma_0 \gamma_1 \gamma_2 \gamma_3$, in four
dimensions. Then, we displace all $\not{n}$ to the right, collect all $n.(p+Q)$ produced by this motion and use them to cancel as many $n.(p+Q)$ in the denominator as possible. At the end, all remaining  $(n.(p+Q))^{-a},a>0$ are substituted by zero. Here, $Q_\mu$ means any vector different from $p_\mu$ (the integration variable), the zero vector included.

Applying this method, we obtain the standard QED answer with the
physical fermion mass: $M_e^2 = M^2 + m^2 .$
\begin{equation}
i \Pi_{^{} 1 \mu \nu}^5 = -e^2 \int d p \frac{\tmop{Tr} \left\{ \gamma_{\mu}
	\not{p} \gamma_{\nu}\gamma_5 \not{p}  + \gamma_{\mu} \not{p} \gamma_{\nu}\gamma_5
	\not{q}  + M_e^2 \gamma_{\mu} \gamma_{\nu} \gamma_5 \right\}}{(p^2
	- M_e^2 + i \varepsilon) ((p + q)^2 - M_e^2 + i \varepsilon)}	\label{axial}
\end{equation}

Using the two dimensional identity:
\[ \gamma^{\mu} \gamma_5 = - \varepsilon^{\mu \alpha} \gamma_{\alpha}
\label{id2} \]
we obtain $i \Pi_{ 1 \mu \nu}^5 =-\epsilon_{\nu\alpha}i \Pi_\mu^\alpha$, which implies
\begin{equation}
		< j^{\mu 5} (q) > = - \varepsilon^{\mu \nu} < j_{\nu} (q) >\label{ac0}
\end{equation}

From the previous subsection, the divergence of the axial current is:
\begin{eqnarray}
	< j^{\mu 5} (q) > q_{\mu} = - \frac{e}{\pi} q^2 \varepsilon^{\mu \nu}
	q_{\mu} A_{\nu} (q) \int_0^1 \frac{d x}{(- q^2 x (1 - x) + M_e^2)} x (1 - x) &\label{ac}
	& 
\end{eqnarray}

Equation (\ref{ac}) agrees with the calculation of the divergence of the axial current obtained in~\cite{AS2} with $M=0$.
In~\cite{jaanom}, we computed the same quantity and obtained the standard anomaly, i.e., Equation (\ref{ac}) with $m=0, M=0$.
We will see below that the different results are a matter of interpretation. In fact,
we can see that a $\tmop{Sim}(2)$ regulator will break the
chiral symmetry by generating a standard fermion mass. So we obtain:
\begin{eqnarray}
	\partial_{\mu} j^{\mu 5} (x) = \frac{e}{2 \pi} \varepsilon_{\mu \nu} F_{\mu
		\nu} + 2 M_e \bar{\psi} \gamma_5 \psi &  & 
\end{eqnarray}

In fact, define $K_\mu$ by:
\begin{equation}
\int d^2 x < \bar{\psi} (x) \gamma_5 \psi (x) > e^{i q x} = K_{\mu}(q)
A_{\mu}(q)	
\end{equation}

We obtain:
\begin{eqnarray}
	K_{\mu} = - (- i e) \int d p \frac{\tmop{Tr} \left\{ \gamma_{\mu} i \left(
		\not{p} + M_e \right) \gamma_5 i \left( \not{p} + \not{q} + M_e \right)
		\right\}}{(p^2 - M_e^2 + i \varepsilon) ((p + q) ^2 - M_e^2 + i
		\varepsilon)} = &  & \nonumber\\
	\frac{M_e}{2 \pi} e \varepsilon_{\mu \alpha} q_{\alpha} \int d x
	\frac{1}{M_e^2 - q^2 x (1 - x) - i \varepsilon} &  & 
\end{eqnarray}

Notice that $K_{\mu} q_{\mu} = 0$, as implied by gauge invariance.

Then:
\begin{eqnarray}
	< j^{\mu 5} (q) > q_{\mu} - 2 M_e K_{\mu} A_{\mu} = \frac{e}{\pi}
	\varepsilon_{\mu \nu} q_{\mu} A_{\nu} (q) &  & 
\end{eqnarray}

The divergence of the chiral current receives two contributions; one
is due to a non-zero mass: $2 M_e \bar{\psi} \gamma_5 \psi$. The other is the
anomaly, which is present even for a massless fermion model with $M=m=0$. The main difference is
that the contribution of the mass is finite and unambiguous. The anomaly term
requires an ultraviolet (UV) regulator, which breaks the chiral symmetry. In a
model with a VSR mass, the infrared regularization introduces a mass-like term
that contributes to the divergence of the chiral current as well as the
standard anomaly.

This observation permits us to understand the results for the anomaly obtained
in~\cite{AS2,jaanom}.

The result of~\cite{AS2} corresponds to considering the non-zero divergence of the
chiral current as the anomaly. Instead, in~\cite{jaanom}, we computed the standard
anomaly, which is present even for zero fermion mass, as a result of the UV
regulator.

\subsection{Anomaly Calculation in Dimensional Regularization}
In this subsection, we will directly compute the expectation value of the chiral current using dimensional regularization, instead of using Equation (\ref{ac0}).
To treat $\gamma^5$, we follow the method of~\cite{mdr}. That is, in any number of
dimensions
\begin{eqnarray}
	\gamma^{5} & =i  \gamma^{0} \gamma^{1} , & \\
	\{ \gamma^{5} , \gamma^{\mu} \} =0. \mu =0,1; & [ \gamma^{5} , \gamma^{\mu}
	] =0, \mu =2,3 \ldots .,d & 
\end{eqnarray}
	
$q_{\mu} ,n^{\mu}$ are two dimensional vectors. 
$p_{\mu}$  is d-dimensional.

After applying the infrared regulator we found in the previous subsection, Equation (\ref{axial})
\begin{equation*}
	i \Pi_{^{} 1 \mu \nu}^5 = -e^2 \int d p \frac{\tmop{Tr} \left\{ \gamma_{\mu}
		\not{p} \gamma_{\nu}\gamma_5 \not{p}  + \gamma_{\mu} \not{p} \gamma_{\nu}\gamma_5
		\not{q}  + M_e^2 \gamma_{\mu} \gamma_{\nu} \gamma_5 \right\}}{(p^2
		- M_e^2 + i \varepsilon) ((p + q)^2 - M_e^2 + i \varepsilon)}	
\end{equation*}

Write $\not{p} = \not{p}_1 + \not{p}_2$, with $p_{1 \mu}, \mu = 0, 1$; $p_{2
	\mu}, \mu = 2, \ldots, d$. Then:

\begin{adjustwidth}{-\extralength}{0cm}
\centering 
\begin{equation}
	i \Pi_{^{} 1 \mu \nu}^5 =- e^2 \int_0^1 d x \int d p \frac{\tmop{Tr} \left\{ \gamma_{\mu}
	\not{p}_1 \gamma_{\nu} \gamma_5 \not{p}_1 + \gamma_{\mu} \not{p}_2
	\gamma_{\nu} \gamma_5 \not{p}_2 + (x^2 - x) \gamma_{\mu} \not{q} \gamma_{\nu}
	\gamma_5 \not{q} + M_e^2 \gamma_{\mu} \gamma_{\nu} \gamma_5 \right\}}{(p^2 -
	M_e^2 + i \varepsilon + q^2 x (1 - x))^2}
\end{equation}
\end{adjustwidth}

Using dimensional regularization, the numerator can be replaced by:
\begin{eqnarray}
	\tmop{Tr} \left( \gamma_{\mu} \not{p}_1 \gamma_{\nu}\gamma_5 \not{p}_1 
	\right) =- \varepsilon_{\nu \alpha} \tmop{Tr} \left( \gamma_{\mu} \not{p}_1
	\gamma_{\alpha} \not{p}_1 \right)  \rightarrow -\varepsilon_{\nu
		\alpha} \left( 4 \frac{1}{2} p_1^2 \eta_{\mu \alpha} - 2 p_1^2 \eta_{\mu
		\alpha} \right) = 0 \\
	\tmop{Tr} \left( \gamma_{\mu} \not{p}_2 \gamma_{\nu}\gamma_5 \not{p}_2 
	\right) = -p_2^2 \tmop{Tr} (\gamma_{\mu}
	\gamma_{\nu}\gamma_5  ) \rightarrow  -p^2 \frac{d - 2}{d} \tmop{Tr} (\gamma_{\mu}
	\gamma_{\nu} \gamma_5) &  & 
\end{eqnarray}

However, 
\[ \frac{d - 2}{d} \int d p \frac{p^2}{(p^2 - M_e^2 + i \varepsilon + q^2 x (1
	- x))^2} = \frac{i}{4 \pi} \]
	
Therefore:
\begin{equation}
	i \Pi_{^{} 1 \mu \nu}^5 =-e^2 \frac{i}{\pi} (q^2 \varepsilon_{\mu \nu} + q_{\mu}
\varepsilon_{\nu \alpha} q^{\alpha}) \int d x \frac{x (1 - x)}{M_e^2 - q^2 x
	(1 - x)}
\end{equation}

So, the divergence of the axial current is
\begin{equation*}
	q_{\mu} < j^{\mu 5} > = \frac{i}{e} i \Pi_1^{ \mu \nu 5} q_{\nu} A_{\mu}
	=  - \frac{e}{2 \pi} q^2 \varepsilon^{\mu \nu} F_{\mu \nu}
	\int d x \frac{x (1 - x)}{M_e^2 - q^2 x (1 - x)} 
\end{equation*}
which coincides with Equation (\ref{ac}).

In the massless case we recover Equation (19.18) of~\cite{Peskin}. If $M = 0$, we have
conservation of chiral symmetry at the classical level in VSR QED, but this
symmetry is broken at the quantum level. We can consider the above expression
as the anomaly, as in~\cite{AS2}, or interpret the anomaly as the one corresponding to
the massless case~\cite{jaanom} and the remaining, being finite and unambiguous, as
the mass contribution to the divergence of the chiral current.

\section{Four-Dimensional Axial Anomaly}\label{sec:5}

We compute:
\[ \int d^{4} x e^{-i r x} <p,q | j^{\mu 5} ( x ) | 0> = ( 2 \pi )^{4} \delta
( -r+p+q ) \varepsilon^{\ast}_{\nu} ( q ) \varepsilon^{\ast}_{\delta} ( p )
i  \Pi^{\mu \nu \delta} \]

\begin{figure}[H]
	\includegraphics[scale=1]{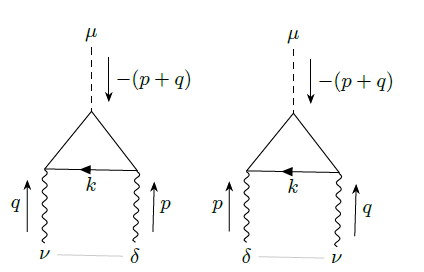}
	\caption{$i\Pi^{ 15\mu \nu \delta}$ contribution to the four-dimensional axial current in VSR QED.}
	\label{fig:4d1}
\end{figure}

\begin{figure}[H]
	\includegraphics[scale=1]{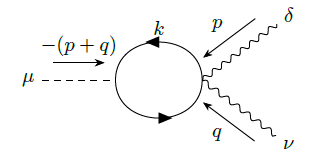}
	\caption{$i\Pi^{2 5\mu \nu \delta}$ contribution to the four-dimensional axial current in VSR QED.}
	\label{fig:4d2}
\end{figure}

\begin{figure}[H]
	\includegraphics[scale=1]{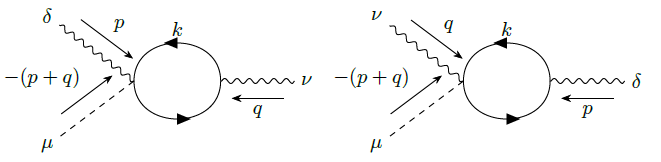}
	\caption{$i\Pi^{3 5\mu \nu \delta}$ contribution to the four-dimensional axial current in VSR QED.}
	\label{fig:4d3}
\end{figure}

There are four topologically distinct graphs, plus permutations of the external legs, that contribute to the axial anomaly in four dimensions
(Figures~\ref{fig:2d2}--\ref{fig:4d3}). The Feynman rules are written in Figure~\ref{Fig: Fey rulesA}. They are fundamental
to satisfying the formal Ward identity for the vector current (charge conservation) as
well as the right computation of the axial anomaly~\cite{jaanom}.
\vspace{-12pt}

\begin{adjustwidth}{-\extralength}{0cm}
\centering 
\begin{eqnarray}
	i  \Pi^{1 5\mu \nu \delta} =- ( -i e )^{2} \int d k  \tmop{Tr}
	\{ \left[ \gamma^{\mu} + \frac{1}{2} n^{\mu} \left( \slashed{n} \right)
	m^{2}   ( n. ( k+q ) )^{-1} ( n. ( k  - p ) )^{-1} \right]
	\gamma^{5}  \frac{i \left( \left( \slashed{k} + \slashed{q} \right) +M-
		\frac{m^{2}}{2}  \frac{\slashed{n}}{n \cdot ( k+q )} \right)}{( k+q )^{2} -M^{2}
		-m^{2} +i \varepsilon} \nonumber \\
	\left[ \gamma^{\nu} + \frac{1}{2} n^{\nu} \left(
	\slashed{n} \right) m^{2}   ( n. ( k+q ) )^{-1} ( n.k )^{-1} \right] \frac{i
		\left( \slashed{k} +M- \frac{m^{2}}{2}  \frac{\slashed{n}}{n \cdot k} \right)}{k^{2}
		-M^{2} -m^{2} +i \varepsilon} \nonumber \\
	\left[ \gamma^{\delta} + \frac{1}{2}
	n^{\delta} \left( \slashed{n} \right) m^{2}   ( n. ( k-p ) )^{-1} ( n.k )^{-1}
	\right] \frac{i \left( \slashed{k} - \slashed{p} +M- \frac{m^{2}}{2} 
		\frac{\slashed{n}}{n \cdot ( k-p )} \right)}{( k-p )^{2} -M^{2} -m^{2} +i
		\varepsilon} \} + ( p, \delta ) \rightarrow ( q, \nu ) &  & 
	\label{4d1}
\end{eqnarray}
\end{adjustwidth}
\vspace{-6pt}

\vspace{-9pt}

\begin{eqnarray}
	i  \Pi^{2 5\mu \nu \delta} = ( -1 ) ( i e )^{2} n^{\delta} n^{\nu} i \int
	d k ( n.k )^{-1} ( n. ( k-p-q ) )^{-1} [ ( n. ( k-q ) )^{-1} + ( n. ( k-p )
	)^{-1} ] \nonumber \\
	\tmop{Tr} \{\frac{1}{2} \slashed{n} m^{2}  \frac{i \left( \slashed{k} -
		\slashed{p} - \slashed{q} +M- \frac{m^{2}}{2}  \frac{\slashed{n}}{n \cdot ( k-p-q )}
		\right)}{( k-p-q )^{2} -M_e^{2}  +i \varepsilon} \left[ \gamma^{\mu} +
	\frac{1}{2} n^{\mu} \left( \slashed{n} \right) m^{2}   ( n.k )^{-1} ( n. ( k-p-q
	) )^{-1} \right] \gamma^{5} &  &  \nonumber\\
	\frac{i \left( \slashed{k} +M- \frac{m^{2}}{2}  \frac{\slashed{n}}{n \cdot k}
		\right)}{k^{2} -M_e^{2} +i \varepsilon} \}   \label{4d2}
\end{eqnarray}
\vspace{-6pt}

\vspace{-12pt}

\begin{adjustwidth}{-\extralength}{0cm}
\centering 
\begin{eqnarray}
	i  \Pi^{3 5\mu \nu \delta}= ( -1 ) ( i e )^{2} n^{\delta} n^{\mu} i \int
	d k ( n.k )^{-1} ( n. ( k-q ) )^{-1} [ ( n. ( k { -} q
	{ -} p ) )^{-1} + ( n. ( k+p ) )^{-1} ] \nonumber \\
	\tmop{Tr} \left\{
	\frac{1}{2} \slashed{n} m^{2} \gamma^{5}  \frac{i \left( \slashed{k} +M-
		\frac{m^{2}}{2}  \frac{\slashed{n}}{n \cdot k} \right)}{k^{2} -M^{2} -m^{2} +i
		\varepsilon} \left[ \gamma^{\nu} + \frac{1}{2} n^{\nu} \left( \slashed{n}
	\right) m^{2}   ( n.k )^{-1} ( n. ( k-q ) )^{-1} \right] \frac{i \left(
		\slashed{k} - \slashed{q} +M- \frac{m^{2}}{2}  \frac{\slashed{n}}{n \cdot ( k-q )}
		\right)}{( k-q )^{2} -M^{2} -m^{2} +i \varepsilon} \right\}\nonumber \\
	+ ( p, \delta )
	\rightarrow ( q, \nu ) &  &  \label{4d3}
\end{eqnarray}
\end{adjustwidth}

\begin{eqnarray}
	i  \Pi^{4 5\mu \nu \delta} =
	( - 1 ) ( i e )^{2} n^{\nu} n^{\mu} n^{\delta} i \int  d
	k \{ \frac{1}{n.k} \frac{1}{n.k} 
	[ \frac{1}{n. ( k+p+q )}
	\frac{1}{n. ( k+p )} + \frac{1}{n. ( k+p+q )} \frac{1}{n. ( k+q )} +\nonumber\\
	\frac{1}{n. ( k-p )} \frac{1}{n. ( k-p-q )} + \frac{1}{n. ( k-q )}
	\frac{1}{n. ( k+p )} + \frac{1}{n. ( k-p )} \frac{1}{n. ( k+q )} +
	\frac{1}{n. ( k-q )} \frac{1}{n. ( k-p-q )} ] \} \nonumber\\
	\tmop{Tr}
	\left\{ \frac{1}{2} \slashed{n} m^{2} \gamma^{5}  [ i ] \frac{i \left( \slashed{k}
		+M- \frac{m^{2}}{2}  \frac{\slashed{n}}{n \cdot ( k )} \right)}{( k )^{2} -M^{2}
		-m^{2} +i \varepsilon} \right\} &  &  \label{4d4}
\end{eqnarray}

\begin{figure}[H]
	\includegraphics[scale=1]{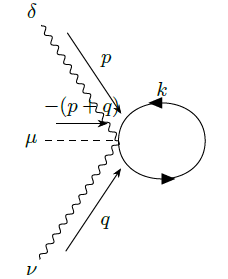}
	\caption{$i\Pi^{4 5\mu \nu \delta}$ contribution to the four-dimensional axial current in VSR QED.}
	\label{fig:4d4}
\end{figure}

However, using the $\bar{n}_{\mu} \rightarrow 0$ limit, we easily see that the
graphs with insertions of non-standard QED vertices, shown in Figures~\ref{fig:4d2}--\ref{fig:4d4}, vanish. In
fact, Figure~\ref{fig:4d2} contains two $n_{\mu}$ outside the integral. $\tmop{Sim}(2)$
symmetry means that there are two $\frac{1}{n. (k + P)}$ inside the integral. That is, the integral is proportional to $\bar{n}
_{\alpha} \bar{n}_{\beta} \rightarrow 0$. The same argument shows that Figures
~\ref{fig:4d3} and \ref{fig:4d4} vanish. 

The standard graphs, shown in Figure~\ref{fig:4d1}, remain.

We have to study the behavior of the
$\bar{n}_{\mu} \rightarrow 0$ limit in the presence of $\gamma_5$. As with the method for the two dimensions, we must use the standard
definition of
$\gamma_5$ in the corresponding dimension from the beginning, $\gamma_5 = i \gamma_0 \gamma_1 \gamma_2 \gamma_3$, in four
dimensions. We displace all $\not{n}$ to the right, collect all $n.(p+Q)$ produced by this motion and use them to cancel as many $n.(p+Q)$ in the denominator as possible. At the end, all remaining  $(n.(p+Q))^{-a},a>0$ are substituted by zero. Here, $Q_\mu$ means any vector different from $p_\mu$ (the integration variable), the zero vector included.

The result agrees with standard QED, with a mass $M_e$:
\begin{eqnarray}
	i \Pi^{15 \mu \nu \delta} = e^2 \int d k \tmop{Tr} \left\{ \gamma^{\mu}
	\gamma_5 i \frac{\not{k} + \not{q} + M_e}{(k + q)^2 - M_e^2 + i \varepsilon}
	\gamma^{\nu} i \frac{\not{k} + M_e}{k^2 - M_e^2 + i \varepsilon}
	\gamma^{\delta} i \frac{\not{k} - \not{p} + M_e}{(k - p)^2 - M_e^2 + i
		\varepsilon} \right\} &  & 
\end{eqnarray}

This is a great simplification.

Therefore, the divergence of the axial vector current has two terms: the same anomaly as in standard QED when  $M=m=0$, plus a mass term if either $M\ne 0$ or $m\ne 0$. If $M=0$ and $m\ne 0$, the axial vector current is conserved classically but is broken at the quantum level by a mass term.

\section{Gross--Neveu Model in VSR}\label{sec:6}
Please see the word of caution at the beginning of section 4.
We follow~\cite{Coleman}. We add a VSR mass $m$ to the fermion.
\begin{eqnarray}
	\mathcal{L}= \bar{\psi}^a \left( i \left( \not{\partial} + + \frac{1}{2}
	\not{n} m^2 (n. \partial)^{- 1} \right) \right) \psi^a + \frac{g_0}{2 N}
	(\bar{\psi}^a \psi^a)^2 &  & 
\end{eqnarray}

The Lagrangian is invariant under the discrete chiral symmetry:
\[ \psi^a \rightarrow \gamma_5 \psi^a, \bar{\psi}^a \rightarrow - \bar{\psi}^a
\gamma_5 \]

The large $N$ limit can be simplified by the introduction of a bosonic field $\sigma$.
\begin{eqnarray}
	\mathcal{L}= \bar{\psi}^a \left( i \left( \not{\partial} + + \frac{1}{2}
	\not{n} m^2 (n. \partial)^{- 1} \right) \right) \psi^a + \frac{g_0}{2 N}
	(\bar{\psi}^a \psi^a)^2 - \frac{N}{2 g_0} \left( \sigma - \frac{g_0}{N}
	\bar{\psi}^a \psi^a \right)^2 \rightarrow &  & \nonumber\\
	\bar{\psi}^a \left( i \left( \not{\partial} + + \frac{1}{2} \not{n} m^2 (n.
	\partial)^{- 1} \right) \right) \psi^a - \frac{N}{2 g_0} \sigma^2 + \sigma
	\bar{\psi}^a \psi^a &  & 
\end{eqnarray}

The generating functional is obtained by integrating over the fermion fields to obtain:
\begin{eqnarray}
	Z = \int d \sigma e^{i \int d x \left( - \frac{N}{2 g_0} \sigma^2 \right) +
		\tmop{tr} \left( \log \left( i \left( \not{\partial} + + \frac{1}{2} \not{n}
		m^2 (n. \partial)^{- 1} + \sigma (x) \right) \right) \right)} &  & 
\end{eqnarray}

To find the vacuum, we just need the effective potential $V_{eff}$. $\sigma$ is independent of $x^\mu$.

\begin{eqnarray}
	- i V_{\tmop{eff}} (\sigma) = - i \frac{N}{2 g_0} \sigma^2 + \tmop{tr}
	\left( \log \left( i \left( \not{\partial} + + \frac{1}{2} \not{n} m^2 (n.
	\partial)^{- 1} + \sigma \right) \right) \right) &  & 
\end{eqnarray}

Thus:
\begin{equation}
	- i V_{\tmop{eff}} (\sigma)  = 
	- i \frac{N}{2 g_0} \sigma^2 - \sum_{n = 1} \frac{N}{2 n} \tmop{Tr} \int d p
	\frac{ (p^2 - m^2)^n}{(p^2 - m^2 + i \varepsilon)^{2 n}} \sigma^{2 n} 
\end{equation}

\begin{figure}[H]
	\includegraphics[scale=1]{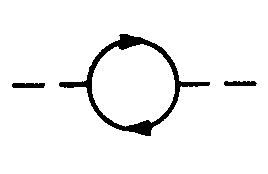}
	\caption{Self energy of the $\sigma$ field at one loop order.}
	\label{fig:2}
\end{figure}

We can see that we do not have infrared divergences in the effective potential. We do need the infrared regulator of Section~\ref{sec:2} if we want to compute the effective action. For instance, Figure~\ref{fig:2} needs the infrared regulator for arbitrary external momentum.
\vspace{-6pt}

We compute the effective potential using dimensional regularization in the minimal subtraction scheme (MS). $\mu$ is the arbitrary scale introduced in dimensional regularization.

The model is renormalizable. The renormalized coupling $g$ in MS is:
$\frac{1}{ g} = \frac{1}{ g_0} - \frac{1}{ \pi \varepsilon},\varepsilon=2-d$

We get:
\begin{equation}
 \frac{1}{N} V_{\tmop{eff}} = \sigma^2 \left( \frac{1}{2 g} - \frac{\Gamma'
	(1)}{4 \pi} - \frac{1}{4 \pi} \right) + \frac{1}{4 \pi} m^2 \log \left(
\frac{\sigma^2}{m^2} + 1 \right) + \frac{\sigma^2}{4 \pi} \log \left(
\frac{\sigma^2 + m^2}{4 \pi \mu^2} \right)	
\end{equation}

To make contact with reference~\cite{Coleman} Equation (2.26), choose
\begin{eqnarray*}
	\log (4 \pi \mu^2) = \log (M^2) + 2 - \Gamma' (1) &  & 
\end{eqnarray*}

We obtain:
\begin{equation}
	\frac{1}{N} V_{\tmop{eff}} = \frac{\sigma^2}{2 g} + \frac{1}{4 \pi} m^2 \log
	\left( \frac{\sigma^2}{m^2} + 1 \right) + \frac{\sigma^2}{4 \pi} \left( \log
	\left( \frac{\sigma^2 + m^2}{M^2} \right) - 3 \right)
\end{equation}

It goes to Coleman's Equation (2.26)  when $m \rightarrow 0$

The ground state $\sigma_0$ is the absolute minimum of $V_{eff}$. It satisfies:
\begin{equation}
	\frac{\sigma_0 \hspace{0.17em} \left( \log \left(
		\frac{{\sigma_0}^2 + m^2}{M^2} \right) - 3 \right)}{2 \pi} +
	\frac{\sigma_0}{2 \pi \left( \frac{\sigma_0^2}{m^2} + 1
		\right)} + \frac{\sigma_0^3}{2 \pi (\sigma_0^2 + m^2)} +
	\frac{\sigma_0}{g} = 0
\end{equation}
i.e., $\sigma_0 = 0$ or

\begin{equation}
	\sigma_0^2 = M^2 e^{2 - \frac{2 \pi}{g}} - m^2
\end{equation}

Notice that now we have the additional condition that $M^2 e^{2 - \frac{2 \pi}{g}} - m^2\ge 0$.
$\sigma_0^2 + m^2$ is a physical quantity. Thus, it satisfies the
homogeneous renormalization group equation:
\begin{eqnarray}
	\left( M \frac{\partial}{\partial M} + \beta (g) \frac{\partial}{\partial g}
	\right) (\sigma_0^2 + m^2) = 0\\
	\beta (g) = - \frac{g^2}{\pi}
\end{eqnarray}

So, in the phase where $M^2 e^{2 - \frac{2 \pi}{g}} - m^2\ge 0$, the model is asymptotically free.
It is easy to verify that if $M^2 e^{2 - \frac{2 \pi}{g}} - m^2\ge 0$, then  $\sigma_0$ is a global minimum of $V_{eff}$.

The running coupling is:
\begin{equation}
	g =\frac{g_0}{1 + \frac{g_0}{\pi} \log \left( \frac{M}{M_0} \right)}	
\end{equation}

The model has two phases:
\begin{enumerate}
	\item $M^2 e^{2 - \frac{2 \pi}{g}} - m^2 \ge 0, \sigma_0 \neq 0$. The discrete
	chiral symmetry is spontaneously broken.
	
	\item $M^2 e^{2 - \frac{2 \pi}{g}} - m^2 < 0, \sigma_0 = 0$. The
	discrete chiral symmetry is preserved.
\end{enumerate}

\section{Conclusions}\label{sec:7}

We have extended the infrared regularization of~\cite{jaren} to include terms with $\gamma_5$. We have checked that the extension preserves $\tmop{Sim}(2)$ and gauge symmetries.

As an application of the infrared regulator, we first consider the two-dimensional anomaly in the Schwinger model. We have explicitly checked that the result is gauge invariant and produces the standard anomaly. Moreover, by computing the mass contribution to the divergence of the axial current, we were able to clarify the different results obtained in previous works~\cite{AS2,jaanom}. In~\cite{AS2}, we computed the divergence of the axial current in the presence of a VSR fermion mass. We find the same answer in the present paper for $M=0$. In~\cite{jaanom}, we computed the standard anomaly produced by the ultraviolet divergences. According to the analysis we have carried out now, the divergence of the axial current receives two contributions: the anomaly, which is present even for massless fermions; and the mass term. If we introduce a VSR fermion mass, the axial vector current is conserved at the classical level, but is violated at the quantum level by the anomaly due to a ultraviolet divergences; and by a mass term. The mass term seems to be unavoidable if we want to preserve $\tmop{Sim}(2)$ symmetry at the quantum level. 

As a second application of the infrared regulator, we computed the triangle anomaly in four-dimensional QED with a VSR fermion mass. Again, the regulator combines the VSR terms in such a way that the final answer is the same as the standard QED one with the physical mass for the fermion $M_e=\sqrt{M^2 +m^2}$. It follows that we obtain the standard anomaly, as in~\cite{jaanom} plus a VSR mass contribution to the divergence of the axial current. 

Finally, we solved the Gross--Neveu(GN) model with a VSR fermion mass. We found that the VSR mass $m$ allows for the existence of two phases. In one of them, the chiral symmetry is broken, as in the standard GN model; in the new phase, the chiral symmetry is unbroken.

These calculations show the ability of the infrared regulator to preserve  $\tmop{Sim}(2)$ and gauge symmetries, as well as to considerably simplify the calculations.

Several possible applications are feasible: the VSR extension of the Nambu--Jona-Lasinio model, considering the implications of VSR masses for fermions in supersymmetric and super string models, and loop computations in the VSRSM, including the three families of quarks and leptons, among others.

\vspace{6pt}

\funding{\hl{~}This research received no external funding.
}

\dataavailability{\hl{~}  No new data were created.
}

\acknowledgments{J.A. acknowledges the partial support of the Institute of
	Physics PUC.}

\conflictsofinterest{\hl{~}The author declare no conflicts of interest.
} 
	

\appendixtitles{yes} 
\appendixstart
\appendix

\section{Feynman Rules}\label{App:a}
To draw the Feynman graphs, we used~\cite{ellis}
\vspace{-6pt}\begin{figure}[H]
	\includegraphics[scale=0.4]{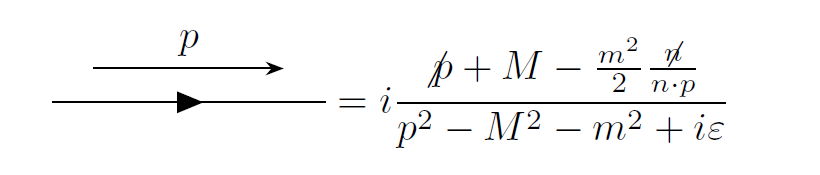} \hspace{6pt}
	\includegraphics[scale=0.4]{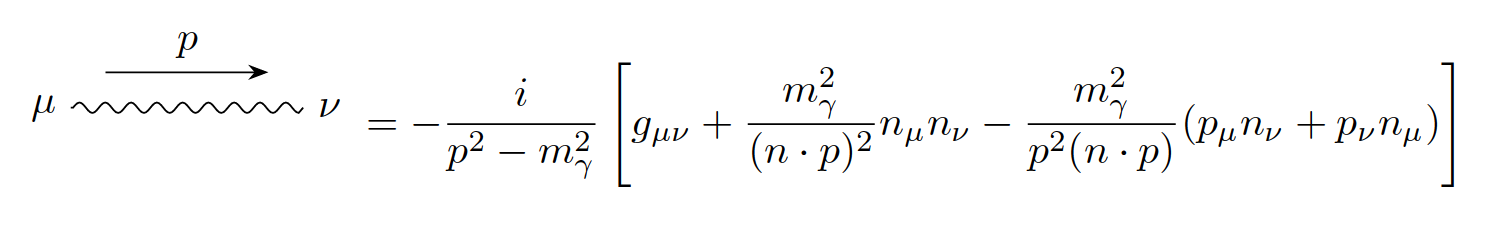} \\
	\includegraphics[scale=0.4]{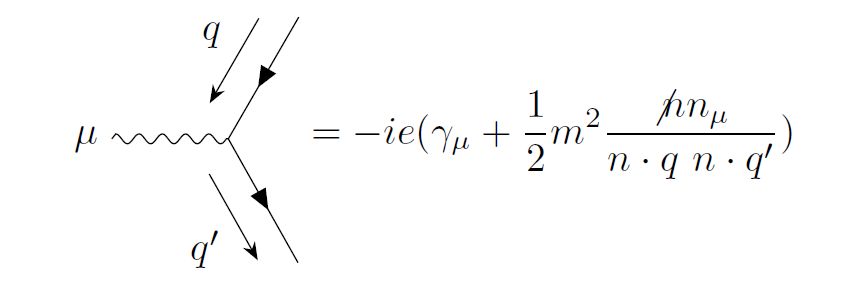} \hspace{6pt}
	\includegraphics[scale=0.4]{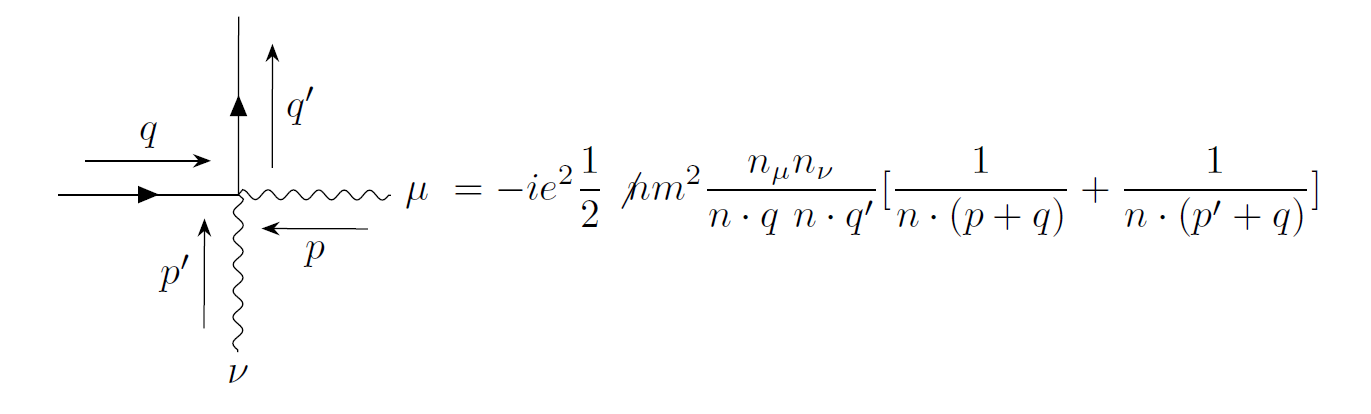}
	\caption{Feynman rules for one loop computations: electron propagator,photon propagator, $A_\mu ee$ and $A_\mu A_\nu ee$ vertex.}
	\label{Fig: Fey rules}
\end{figure}
\vspace{-6pt}
\begin{figure}[H]
	\includegraphics[scale=0.4]{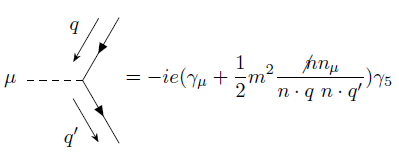}
	\includegraphics[scale=0.4]{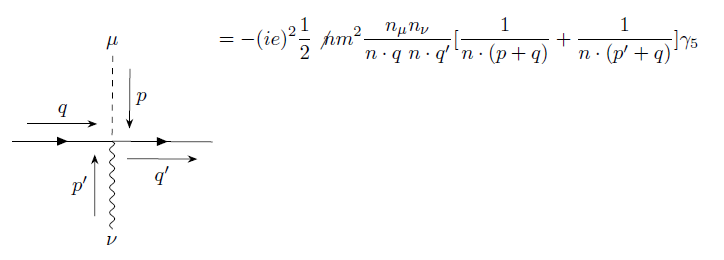}
	\includegraphics[scale=0.4]{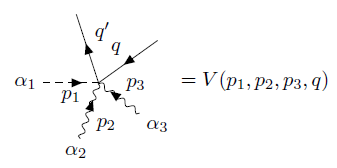}
	\caption{Feynman rules for one loop computations: axial-e-e
		vertex, axial-$A_{\nu}$-e-e vertex and axial-$A_{\alpha_{2}}-A_{\alpha_{3}}$-e-e vertex.}
	\label{Fig: Fey rulesA}
\end{figure}	
\vspace{-12pt}

\begin{eqnarray*}
	V ( p_{1} ,p_{2} ,p_{3} ,q ) =i (ie)^{3} \frac{m^{2}}{2} \slashed{n}
	n^{\alpha_{1}} n^{\alpha_{2}} n^{\alpha_{3}}  \frac{1}{n. (q+p_{1} +p_{2}
		+p_{3} )}\\
	( \frac{1}{n. (q+p_{1} +p_{2} )}  \frac{1}{n. (q+p_{1} )} + \frac{1}{n.
		(q+p_{1} +p_{2} )}  \frac{1}{n. (q+p_{2} )} +\\
	\frac{1}{n. (q+p_{3} +p_{2} )}  \frac{1}{n. (q+p_{3} )} + \frac{1}{n.
		(q+p_{1} +p_{3} )}  \frac{1}{n. (q+p_{1} )} +\\
	\frac{1}{n. (q+p_{2} +p_{3} )}  \frac{1}{n. (q+p_{2} )} + \frac{1}{n.
		(q+p_{3} +p_{1} )}  \frac{1}{n. (q+p_{3} )} ) \gamma^{5}
\end{eqnarray*}
	
\section{Method of Traces}\label{App:b}

To obviate the looseness in the method, we define a basis of gamma functions and develop any matrix function $A$ in this basis:
\begin{eqnarray*}
	1, \gamma_{\mu}, \sigma_{\mu \nu}, \ldots . &  & \\
	A = trace(A) 1 + trace(A\gamma_\alpha)
	\gamma_{\alpha} + \ldots . &  & 
\end{eqnarray*}

All traces are functions of $n_\mu,\bar{n}_\mu$ and external momenta. To obtain the limit  $\bar{n}_\mu \rightarrow 0$, we follow the ensuing steps. First, in all monomials, use the identities $n.n=\not{n}.\not{n}=0$, $n.\not{n}=1$. Afterwards, put $\bar{n}_\mu=0$ everywhere. 

In addition, the identity used in Equation (7.65) of~\cite{Peskin} will be enforced in each trace. Due to the linear and cyclic properties of the trace, the demonstration of the Ward--Takahashi identity will go through. Then, the trace method always lead to a gauge invariant, $\tmop{Sim}(2)$ invariant result.

We have checked that in VSR QED, the trace method gives the same results as the method explained in Section~\ref{sec:2}. 

\begin{adjustwidth}{-\extralength}{0cm}

\reftitle{References}

\PublishersNote{}
\end{adjustwidth}

\begin{thebibliography}{999}
	\bibitem{Langacker}Langacker, P. \emph{The Standard Model and Beyond}; CRC
	Press: A Taylor and Francis Group: Boca Raton, FL, USA; 2010.  

	
	\bibitem{mohapatra}Mohapatra, R. \emph{Unification and Supersymmetry: The
	Frontiers of Quark-Lepton Physics}, 3rd ed.; Springer:  New York, New York,USA 2003.
	
	\bibitem{PA} Abraham, J. et al. [Pierre Auger Collaboration] Observation of the Suppression of the Flux of Cosmic Rays above 4 $\times$ 10$^{19}$ eV. \emph{Phys. Rev. Lett.} \textbf{2008}, \emph{101}, 061101.
	\bibitem{AC}Amelino-Camelia, G. Quantum-Spacetime Phenomenology. \emph{Living Rev. Relativ.} \textbf{2013}, \emph{16},  5. https://doi.org/10.12942/lrr-2013-5.
	
	\bibitem{JLM} Jacobson, T.; Liberati, S.; Mattingly, D. Astrophysical Bounds on Planck Suppressed Lorentz Violation. \emph{Lect. Notes Phys.} \textbf{2005}, \emph{669}, 101.
	
	\bibitem{MP}Myers, R.C.; Pospelov, M. Ultraviolet Modifications of Dispersion Relations in Effective Field Theory. \emph{Phys. Rev. Lett.} \textbf{2003}, \emph{90}, 211601.
	
	\bibitem{AA}Andrianov, A.A.; Giacconi, P.; Soldati, R. Lorentz and CPT violations from Chern-Simons modifications of QED.\emph{JHEP} \textbf{2002}, \emph{2002}, 030.	
	\bibitem{CK} Colladay, D.; Kostelecky, V.A. Lorentz-violating extension of the standard model \emph{Phys.
	Rev. D} \textbf{1998}, \emph{58}, 116002.

	
	\bibitem{CG1}Cohen, A.G.; Glashow, S.L. Very special relativity.
	\emph{Phys. Rev. Lett.} \textbf{2006}, \emph{97}, 021601.
	
	\bibitem{CG2}Cohen, A.; Glashow, S. A Lorentz-Violating Origin of
	Neutrino Mass? \emph{arXiv} \textbf{2006}, arXiv:hep-ph/0605036.
	
	\bibitem{CZ} Cohen, A.G.; Freedman, D.Z. \emph{JHEP} \textbf{2007}, \emph{0707}, 039.
	\bibitem{Vo} Vohanka, J. \emph{Phys. Rev. D} \textbf{2012}, \emph{85}, 105009.
		\bibitem{GG} Gibbons, G.W.; Gomis, J.; Pope, C.N. General Very Special Relativity in Finsler Geometry. \emph{Phys. Rev. D} \textbf{2007}, \emph{76}, 081701.
	\bibitem{Mu} Muck, W. \emph{Phys. Lett. B} \textbf{2008}, \emph{670}, 95.
	
	\bibitem{ST}Sheikh-Jabbari, M.M.; Tureanu, A. Realization of Cohen-Glashow Very Special Relativity on Noncommutative Space-Time. \emph{Phys. Rev. Lett.} \textbf{2008}, \emph{101}, 261601.
	\bibitem{Das} Das, S.; Ghosh, S.; Mignemi, S. Non-commutative spacetime in very special relativity.  \emph{Phys. Lett. A} \textbf{2011}, \emph{375}, 3237.
		
	\bibitem{AV}Alvarez, E.; Vidal, R. \emph{Phys. Rev. D} \textbf{2008}, \emph{77}, 127702.
	
	\bibitem{AH}Ahluwalia, D.V.; Horvath, S.P. \emph{JHEP} \textbf{2010}, \emph{1011}, 078.
	
	\bibitem{CL} Chang, Z.; Li, M.-H.; Li, X.; Wang, S. Cosmological model with local symmetry of very special relativity and constraints on it from supernovae. \emph{Eur. Phys. J. C} \textbf{2013}, \emph{73}, 2459.
	
	\bibitem{Cheon}Cheon, S.; Lee, C.; Lee, S. SIM(2)-invariant modifications of electrodynamic theory. \emph{Phys. Lett. B} \textbf{2009}, \emph{679}, 73.
	
	\bibitem{AR} Alfaro, J.; Rivelles, V. Non Abelian Fields in Very Special Relativity. \emph{Phys. Rev. D} \textbf{2013}, \emph{88}, 085023.
	
	\bibitem{ja1}Alfaro, J.; Gonz{\'a}lez, P.; {\'A}vila, R. Electroweak standard model with very special relativity. \emph{Phys. Rev. D} \textbf{2015}, \emph{91}, 105007; Addendum in \emph{Phys. Rev. D} \textbf{2015}, \emph{91}, 129904.
	
	\bibitem{jaren} Alfaro, J. Renormalization of Very Special Relativity gauge theories.\emph{ J. High Energy Phys.} \textbf{2023}, \emph{2023}, 3. \url{https://doi.org/10.1007/JHEP06(2023)003}.

	\bibitem{jalbl} Alfaro, J. Light-Light scattering in Very Special Relativity Quantum Electrodynamics and Cosmic Anisotropies, Phys.Lett.B 858 (2024) 139021 • e-Print: 2406.04381 [physics.gen-ph]
	\bibitem{AS2}Alfaro, J.; Soto, A. Schwinger model a la Very Special Relativity. \emph{Phys. Lett. B} \textbf{2019}, \emph{797}, 134923.
	
	\bibitem{jaanom} Alfaro, J. Axial anomaly in very special relativity. \emph{Phys. Rev. D} \textbf{2021}, \emph{103}, 075011
\bibitem{ML}Mandelstam, S. Light-cone superspace and the ultraviolet finiteness of the $N$=4 model. \emph{Nucl. Phys. B} \textbf{1983}, \emph{213}, 149.
\bibitem{Lei} Leibbrandt, G. Light-cone gauge in Yang-Mills theory. \emph{Phys. Rev. D} \textbf{1984}, \emph{29}, 1699.
	
		\bibitem{AML}Alfaro, J. Mandelstam-Leibbrandt prescription. \emph{Phys. Rev. D} \textbf{2016}, \emph{93}, 065033; Erratum in \textbf{2016}, \emph{94}, 049901.
	
	\bibitem{Peskin} Peskin, M.E.; Schroeder, D.V. \emph{An Introduction to Quantum Field Theory};  Perseus Books: Reading, Ma, USA, 1995; Chapter
 19.1. The convention is $\epsilon^{01}=+1$.
	
	\bibitem{mdr} ’t Hooft, G.; Veltman, M.J.G. Regularization and Renormalization of Gauge Fields.
	\emph{Nucl. Phys. B} \textbf{1972}, \emph{44}, 189--213.
	
	\bibitem{Coleman} Coleman, S. \emph{Aspects of Symmetry}; Cambridge University Press: Cambridge, UK, 1985.
	
	\bibitem{ellis} Ellis, J. TikZ-Feynman: Feynman diagrams with TikZ. \emph{arXiv} \textbf{2016}, arXiv:1601.05437.
\end{thebibliography}
\end{document}